\documentclass[amsmath,amssymb,twocolumn,nofootinbib]{revtex4-1}
\begin{document}
\title{Induced Gravity and Quantum Cosmology}
\author{Alexander~Yu.~Kamenshchik}
\email{kamenshchik@bo.infn.it}
\affiliation{Dipartimento di Fisica e Astronomia, Universit\`a di Bologna\\ and INFN,  Via Irnerio 46, 40126 Bologna,
Italy,\\
L.D. Landau Institute for Theoretical Physics of the Russian
Academy of Sciences,\\
 Kosygin str. 2, 119334 Moscow, Russia}
\author{Alessandro~Tronconi}  
\email{tronconi@bo.infn.it}
\affiliation{Dipartimento di Fisica e Astronomia, Universit\`a di Bologna\\ and INFN, Via Irnerio 46, 40126 Bologna,
Italy}
\author{Giovanni~Venturi}
\email{giovanni.venturi@bo.infn.it}
\affiliation{Dipartimento di Fisica e Astronomia, Universit\`a di Bologna\\ and INFN,  Via Irnerio 46, 40126 Bologna,
Italy}

    \newcommand{\be}[1]{\begin{equation}\label{#1}}
    \newcommand{\ba}[1]{\begin{eqnarray}\label{#1}}
    \newcommand{\ep}[1]{\epsilon_{#1}}
    \newcommand{\ga}[1]{\gamma_{#1}}
    \newcommand{\de}[1]{\delta_{#1}}
    \newcommand{\hdp}{\left(H^{2}\right)'}
    \newcommand{\hd}{H^{2}}
    \newcommand{\rd}{{\rm d}}
    \newcommand{\ds}{{\delta\sigma}}
    \newcommand{\re}{{\rm e}}
    \newcommand{\pa}[1]{\left(#1\right)}
    \newcommand{\paq}[1]{\left[#1\right]}
    \newcommand{\pag}[1]{\left\{#1\right\}}
    \newcommand{\av}[1]{\langle#1\rangle}
    \newcommand{\M}{{\rm M_{\rm P}}}
     \newcommand{\z}{{\zeta}}
    \def\ee{\end{equation}}
    \def\ea{\end{eqnarray}}
    \def\ave{{\rm ave}}
    \def\svar{{\rm svar}}
    \def\half{{1\over 2}}
    \def\x{\mbox{\bf x}}
    \def\k{\mbox{\bf k}}
    \def\p{\mbox{\bf p}}
    \def\R{\mbox{\bf R}}
    \def\aPhi{\langle\Phi\rangle}
    \def\aPsi{\langle\Psi\rangle}
    \def\alphat{\tilde{\alpha}}
    \def\betat{\tilde{\beta}}
    \def\aphi{\phi_0}
    \newcommand{\Pk}{\Phi_{\bf k}}
    \newcommand{\pk}{\delta \phi_{\bf k}}
    \newcommand{\ck}{\delta \chi_{\bf k}}
    \newcommand{\w}{\widetilde}
    \renewcommand{\labelenumi}{\Alph{enumi}}
    \def\ktchi{|\tilde \chi\rangle}
    \def\btchi{\langle\tilde \chi|}
    \def\kschi{|\chi_{s}\rangle}
    \def\bschi{\langle\chi_{s}|}

\begin{abstract}
We study the Wheeler-DeWitt equation for a class of induced gravity models in the minisuperspace approximation. In such models a scalar field nonminimally coupled to gravity determines the effective Newton's constant. For simplicity our analysis is limited to power-law potentials for the scalar field which have exact classical solutions. We show that these models have exact solutions also when quantised. Finally the Einstein Frame form of these solutions is obtained and a classical-quantum correspondence is found.
Realistic induced gravity models also must include a symmetry breaking term which is needed in order to obtain a gravitational constant, successful inflation and a subsequent standard cosmological evolution. Nonetheless the potentials considered are important as they may describe the inflationary phase when the symmetry breaking part of the potential is negligible.  
\end{abstract}
\maketitle
\section{Introduction}
Since the introduction of inflation 
\cite{inflation}, many models have been suggested in order to achieve it.
Subsequently the wealth of data obtained from the Planck survey \cite{Planck} has put severe constraints on the space of the diverse models. In particular the fact that scalar and tensor perturbations \cite{pert} are nearly scale invariant severely restricts such models. Indeed the models which currently best fit the data are the Starobinsky $R+R^2$ \cite{Star-model} model and non-minimal Higgs inflation \cite{higgsinf} (of course one may also consider combinations of the two \cite{vanzo}). The reason for this is that in both cases during inflation, or if we wish in a regime of high curvature, they are almost scale invariant. The two models are related insofar there exists a general equivalence between $f(R)$ gravity (of which Starobinsky inflation is part) and scalar tensor theories (of which Higgs inflation is part). Moreover such theories can be reformulated, through a field redefinition and a conformal transformation of the metric tensor, in terms of General Relativity (GR) and a minimally coupled scalar field. Such a transformation is called the transition from the Jordan Frame (JF) to the Einstein Frame (EF) \cite{Wagoner}. The complete physical equivalence of the two frames, beyond the classical level (and eventually adding quantum inflationary fluctuations to the classical homogeneous background), is still not clear \cite{debate} but it is commonly exploited to calculate the primordial inflationary spectra or to describe the crossing of classical cosmic singularities \cite{KPTVV-sing}.\\

The fact that inflation is associated with intense gravitational fields makes one suspect that quantum cosmology leads to effects on the observed scalar and tensor perturbation spectra. Hence we feel that it would be of interest to study some simple induced gravity models \cite{induced} in the context of quantum gravity, in particular the Wheeler--DeWitt (WdW) equation \cite{DeWitt}, with the aim of finding some, preferably exact, solutions for quantum homogeneous scalar field minisuperspace models. Induced gravity models are a subset of the more general class called scalar-tensor theories (to which Higgs inflation also belongs) and are natural generalisation of GR especially in the presence of large quantum effects which ``induce'' nontrivial coupling between the scalar field and the gravitational sector \cite{tronco}.\\
In particular we shall study general actions of the form
\be{lag0}
S=\int d^4 x\sqrt{-g}\pa{-\frac{1}{2}g^{\mu\nu}\partial_{\mu}\sigma\partial_{\nu}\sigma-V(\sigma)+U(\sigma)R}
\ee
involving a homogeneous scalar field $\sigma$ and a minisuperspace metric
\be{metric}
\rd s^2=
-N(t)^2\rd t^2+a(t)^2\pa{\rd r^2+r^2\rd \Omega^2},
\ee
where $N(t)$ is the lapse function and $a(t)$ is the scale factor.
One may search for a static classical solution to the above by solving the (static) equation of motion 
\be{cleq}
\frac{\rd U}{\rd \sigma} R=\frac{\rd V}{\rd \sigma}
\ee
and a static solution $U=U(\sigma_0)\equiv U_0$, $R=R_0$ should also solve the Einstein equation
\be{clein}
 U_0R_0=2V(\sigma_0).
\ee
Further one has the following requirement \cite{faraoni} for the stability of the solution
\be{stability}
\frac{R_0\paq{U_0''+\frac{\pa{U_0'}^{2}}{U_0}}-\left.\frac{\rd^2 V}{\rd \sigma^2}\right|_{\sigma=\sigma_0}}{1+3\frac{\pa{U_0'}^2}{U_0}}\le 0.
\ee
The use of the global scale invariant potential $V=\lambda \sigma^4$ \cite{cooper} is particularly attractive since it describes a scale invariant inflationary phase which ends in a scale dependent fixed point and is related to the previously mentioned phenomenologically successful models. The final scale dependent fixed point can be thought to arise through the presence of a condensate or the presence of symmetry breaking quantum corrections \cite{cerioni}. We shall not address such points here since we expect them to be important in the end of the inflationary phase when quantum gravitational effects are presumably not significant.\\
The article is organised as follows. In the second Section we illustrate the Hamiltonian formalism for a non minimally coupled scalar field and we then obtain the WdW equation for induced gravity. In Section III we calculate the solutions to the WdW by 3 different approaches, two of which lead to the same class of solutions. In Section IV we formulate the theory in the EF, we show its equivalence to the JF and we illustrate the correspondence between the quantum and the classical solutions. Finally in Section V we present the conclusions. 
\section{Induced gravity}
For the Friedmann flat universe with the metric (\ref{metric}) the Lagrangian (\ref{lag0}) becomes 
\begin{equation}
\mathcal{L} = \frac{6Ua\dot{a}^2}{N}+\frac{6\dot{a}a^2\dot{\sigma}U'}{N}-\frac{a^3\dot{\sigma}^2}{2N}+NVa^3,
\label{Lagrange}
\end{equation}
where a ``dot'' means the derivative with respect to the time parameter $t$ and ``prime" is the derivative with respect to the field $\sigma$. 
The conjugate momenta are
\begin{equation}
p_a=\frac{12\dot{a}aU}{N}+\frac{6a^2U'\dot{\sigma}}{N},
\label{p-a}
\end{equation}
\begin{equation}
p_{\sigma}=\frac{6\dot{a}a^2U'}{N}-\frac{a^3\dot{\sigma}}{N}.
\label{p-sigma}
\end{equation}
On inverting,
\begin{equation}
\dot{a} = \frac{Np_a}{12a(U+3U'^2)}+\frac{Np_{\sigma}U'}{2a^2(U+3U'^2)},
\label{a-dot}
\end{equation}
\begin{equation}
\dot{\sigma} = \frac{Np_aU'}{2a^2(U+3U'^2)} - \frac{Np_{\sigma}U}{a^3(U+3U'^2)}.
\label{sigma-dot}
\end{equation}
Correspondingly, the Hamiltonian has the structure 
\begin{equation}
H = N{\cal H},
\label{Hamilton}
\end{equation}
where the super-Hamiltonian constraint is 
\begin{eqnarray}
&&{\cal H} = \frac{p_a^2}{24a(U+3U'^2)}+\frac{p_ap_{\sigma}U'}{2a^2(U+3U'^2)}\nonumber \\
&&-\frac{p_{\sigma}^2U}{2a^3(U+3U'^2)}-Va^3=0.
\label{super}
\end{eqnarray}
Henceforth we shall restrict our study to the induced gravity case with 
\begin{equation}
U(\sigma) = \gamma \frac{\sigma^2}{2},\; U'=\gamma\sigma, \ U+3U'^2 = \frac12\gamma(1+6\gamma)\sigma^2
\label{induced}
\end{equation}
and 
\begin{equation}
{\cal H} = \frac{p_a^2}{12\gamma(1+6\gamma)a\sigma^2}+\frac{p_a p_{\sigma}}{(1+6\gamma)a^2\sigma}-\frac{p_{\sigma}^2}{2a^3(1+6\gamma)}-Va^3.
\label{super-ind}
\end{equation}
The quantum realisation of the momentum operators in the coordinate representation ($\hbar=1$) is 
\begin{eqnarray}
&&p_a = -i\frac{\partial}{\partial a},\nonumber \\
&&p_{\sigma} = -i\frac{\partial}{\partial \sigma}
\label{momenta}
\end{eqnarray}
and the WdW equation takes the following form
\begin{eqnarray}
&&\left[\frac{1}{12\gamma}\frac{\partial^2}{\partial \pa{\ln a}^2}+\frac{\partial^2}{\partial \ln a\; \partial\ln \sigma}
-\frac{1}{2}\frac{\partial^2}{\partial \pa{\ln\sigma}^2}\right.\nonumber \\
&&\left.+(1+6\gamma)a^6\sigma^2V(\sigma)\right]\Psi(a,\sigma)=0.
\label{WdW}
\end{eqnarray}
Let us note that a particular ordering choice has been made in order to promote the classical constraint (\ref{super}) to a quantum WdW equation (\ref{WdW}) and we omit dimensional factors when irrelevant. For the induced gravity case, in contrast with the case with a minimally coupled scalar field, both the scale factor $a$ and the homogeneous field $\sigma$ kinetic terms involve ordering ambiguities.\\

For simplicity we shall restrict the analysis to the class of power-law potentials
\be{plpot}
V=\lambda M^{4-n}\sigma^n,
\ee
where $\lambda$ is a dimensionless coupling constant and $M$ is an arbitrary energy scale, which, at the classical level, admit exact analytical solutions of the form 
\be{clsol1}
\sigma=\sigma_0\pa{\frac{a_0}{a}}^{\frac{\gamma\pa{n-4}}{1+\gamma\pa{n+2}}},\;H=H_0\pa{\frac{a_0}{a}}^{\frac{\gamma\pa{n-2}\pa{n-4}}{2\paq{1+\gamma\pa{n+2}}}},
\ee
where $\sigma_0$, $H_0$ and $a_0$ are integration constants. The above solutions are attractors for a larger set of solutions of the classical equations (with $n>0$ and $\gamma>0$) and can be mapped into the well known Einstein frame solutions for power-law inflation driven by an exponential potential \cite{exp}. In the absence of the scalar field potential ($V=0$) the following exact solutions also exist
\begin{eqnarray}
&&\sigma=\sigma_0\pa{\frac{a}{a_0}}^{6\gamma\pm\sqrt{6\gamma\pa{1+6\gamma}}},\nonumber \\
&&H=H_0\pa{\frac{a}{a_0}}^{-3-2\paq{6\gamma\pm\sqrt{6\gamma\pa{1+6\gamma}}}}.
\label{clsol23}
\end{eqnarray}
\section{Quantum Solutions}
Let us look for a solution of the WdW equation in the following form:
\begin{equation}
\Psi(a,\sigma) =  \pa{\frac{a}{a_0}}^{\nu}\chi(x),
\label{factor}
\end{equation} 
where the new variable $x$ is 
\begin{equation}
x \equiv a^3\sigma^{\frac{n+2}{2}}.
\label{x}
\end{equation}
Then, Eq. (\ref{WdW}) becomes
\begin{widetext}
\be{wdwans1}
\paq{1-\frac{\gamma^2\pa{n-4}^2}{\Gamma^2}}\frac{\rd^2\chi}{\rd \pa{\ln x}^2}+\frac{2}{3}\nu\paq{1+\frac{6\gamma^2\pa{n-4}}{\Gamma^2}}\frac{\rd\chi}{\rd \ln x}+W_1(x)\chi=0,
\ee
\end{widetext}
where $\Gamma\equiv\sqrt{6\gamma\pa{1+6\gamma}}$ and
\be{Wdef}
W_1(x)\equiv \paq{\frac{2\gamma\nu^2}{3\Gamma^2}+ \frac{4}{3}\gamma\,\lambda\, M^{4-n}x^2}.
\ee
The general solution of (\ref{wdwans1}) can be written in terms of Bessel functions in the form
\begin{equation}
\chi(x) = x^{q}\paq{c_1J_{r}\left(A\,x\right)+c_2Y_{r}\left(A\,x\right)}
\label{eq-chi2}
\end{equation}
with
\be{defans1a}
A\equiv\sqrt{\frac{4\gamma\,\Gamma^2 \lambda\,M^{4-n}}{3\paq{\Gamma^2-\gamma^2\pa{n-4}^2}}},
\ee
\be{defans1b}
r\equiv \sqrt{q^2-\frac{2\gamma\nu^2}{3\paq{\Gamma^2-\gamma^2\pa{n-4}^2}}}
\ee
and
\be{defans1c}
q\equiv-\frac\nu3\frac{\Gamma^2+6\gamma^2\pa{n-4}}{\Gamma^2-\gamma^2\pa{n-4}^2}.
\ee
An analogous procedure can be followed starting from the ansatz
\begin{equation}
\Psi(a,\sigma) =  \pa{\frac{\sigma}{\sigma_0}}^{\mu}\bar \chi(x),
\label{factor2}
\end{equation} 
finally leading to the equation 
\begin{eqnarray}
&&\paq{1-\frac{\gamma^2\pa{n-4}^2}{\Gamma^2}}\frac{\rd^2\bar \chi}{\rd \pa{\ln x}^2}\nonumber \\
&&-4\mu\frac{\gamma^2\pa{n-4}}{\Gamma^2}\frac{\rd\bar \chi}{\rd \ln x}+W_2(x)\bar\chi=0,
\label{wdwans2}
\end{eqnarray}
where 
\be{Wdef}
W_2(x)\equiv \paq{-\frac{4\gamma^2\mu^2}{\Gamma^2}+ \frac{4}{3}\gamma\,\lambda\, M^{4-n}x^2}. 
\ee
This last equation admits the following general solution in terms of Bessel functions
\begin{equation}
\bar \chi(x) = x^{p}\paq{c_1J_{s}\left(A\,x\right)+c_2Y_{s}\left(A\,x\right)}
\label{eq-chi2b}
\end{equation}
with 
\be{defans1b}
s\equiv \sqrt{p^2+\frac{4\mu^2\gamma^2}{\Gamma^2-\gamma^2\pa{n-4}^2}}
\ee
and
\be{defans1c}
p\equiv\frac{2\mu\gamma^2\pa{n-4}}{\Gamma^2-\gamma^2\pa{n-4}^2}.
\ee
Let us note that the solutions obtained from the different ansatzes (\ref{factor}) and (\ref{factor2}) are indeed the same as one may easily verify by the following substitution:
\be{solmap}
\mu=-\nu\pa{\frac{n+2}{6}},\;\bar \chi=x^{\nu/3}\chi. 
\ee
As far as we know a second (distinct) possible set of solutions can be found. Consider the change of variable $(a,\sigma)\rightarrow (u_{\pm},v_{\pm})$ with
\be{chvaru}
u_{\pm}
=a^{3\pa{1\pm\frac{\pa{4-n}\gamma}{\Gamma}}}\sigma^{\frac{n+2}{2}}\paq{\sigma^{{\pm \frac{3\paq{\pa{n+2}\gamma+1}}{\Gamma}}}+\sigma^{{\mp\frac{3\paq{\pa{n+2}\gamma+1}}{\Gamma}}}},
\ee
\be{chvarv}
v_{\pm}
=a^{3\pa{1\pm\frac{\pa{4-n}\gamma}{\Gamma}}}\sigma^{\frac{n+2}{2}}\paq{\sigma^{{\pm \frac{3\paq{\pa{n+2}\gamma}+1}{\Gamma}}}-\sigma^{{\mp\frac{3\paq{\pa{n+2}\gamma+1}}{\Gamma}}}}.
\ee
The WdW equation then takes the form of the following, massive, 2 dimensional Klein Gordon equation
\be{WdWKG}
\pa{\partial_{u_{\pm}}^2-\partial_{v_{\pm}}^2+B}\tilde \Psi(u_{\pm},v_{\pm})=0,
\ee
where $B=\frac{\gamma}{3}\frac{\Gamma^2\lambda\,M^{4-n}}{\Gamma^2-\pa{n-4}^2\gamma^2}$.
Starting from the ansatz $\tilde \Psi=\exp\pa{i q v_{\pm}} \rho(u_{\pm})$ one then finds 
the solution given by
\begin{widetext}
\be{wavesol}
\tilde \Psi_\pm=c_1 \exp\paq{i\pa{q v_{\pm}+\sqrt{q^2+B}u_{\pm}}}+c_2 \exp\paq{i\pa{q v_{\pm}-\sqrt{q^2+B}u_{\pm}}},
\ee
\end{widetext}
where $q$ is a free parameter. Let us note that for $n=4$ the solutions $\Psi_+$ and $\Psi_-$ coincide (simply exchange $q\rightarrow-q$) while for $n\neq 4$ the two solutions given by (\ref{wavesol}) are distinct.\\
So far we have illustrated 3 different methods leading to different sets of solution of the original equation (\ref{WdW}) (although the first two are related).
 
Even if the methods are quite straightforward the form of the solutions found is still rather cumbersome. A rather natural interpretation of the solution 
(\ref{factor}) with $\nu=0$ (considering $n=4$ for simplicity) can be obtained by evaluating the effect on $\chi(x)$ of the operator $\hat h$ defined as the quantum counterpart of the Hubble parameter $h_{\rm cl}$. In terms of the momenta $p_a$ and $p_\sigma$ defined by (\ref{p-a})-(\ref{p-sigma}) one has:
\be{defhop}
\frac{\dot a}{a}\equiv h_{\rm cl}=\frac{a\,p_a+6\gamma \sigma\,p_\sigma}{\Gamma^2}\frac{\sigma}{x}
\ee
and, after quantisation
\be{hquant}
\hat h\chi(x)=-i\frac{\sigma}{2\gamma}\frac{\rd \chi}{\rd x},
\ee
where $\chi$ satisfies Eq. (\ref{wdwans1}) which, for $\nu=0$ and $n=4$, has the simple form
\be{wdwansn4}
\frac{\rd^2\chi}{\rd \pa{\ln x}^2}+\frac{4}{3}\gamma\lambda x^2\chi=0.
\ee
For $x$ large  Eq. (\ref{wdwansn4}) has solutions of the form 
\be{wdwansn4sol}
\chi\simeq \re^{\pm i\sqrt{\frac{4}{3}\gamma\lambda}}
\ee
and correspondingly
\be{hquant2}
\hat h\chi(x)=\pm \sqrt{\frac{\lambda}{3\gamma}}\sigma\,\chi(x)=h_{\rm cl} \chi(x),
\ee
where $h_{\rm cl}$ is the value of the Hubble parameter corresponding to the classical solution with $\dot \sigma=0$. We can then conclude that the solution with $\nu=0$ corresponds to classical static configurations.\\

In order to get a better understanding of such solutions, in the next section we shall transform them to the EF form where the degrees of freedom do not mix as in the JF and we restrict our attention to the physically more interesting case with $n=4$. For such a case, the original Lagrangian (\ref{Lagrange}) become scale invariant, a condition needed in order to generate the correct inflationary spectra.


\section{Einstein Frame Transition}
As we already discussed in the Introduction, at the classical level there is a well known redefinition of the dynamical degrees of freedom in the inflaton-gravity action (\ref{lag0}), called transformation from the JF to the EF. This transformation maps the original metric and the scalar field $\pa{g_{\mu\nu},\sigma}$ into a new metric and a redefined scalar field $\pa{\tilde g_{\mu\nu},\phi}$ with an action given by the standard Einstein Hilbert term and a minimally coupled scalar field:
\be{EFaction}
S_E=\int \rd^4x \sqrt{-\tilde g}\left(\frac{\M^2}{2}R-\frac12\tilde g^{\mu\nu}\phi_{,\mu}\phi_{,\nu}+W(\phi)\right).
\ee
This transformation corresponds to a conformal rescaling of the metric
\be{confmet}
g_{\mu\nu}=\frac{\M^2}{2U(\sigma)}\tilde g_{\mu\nu}
\ee
and the following redefinition of the scalar field 
\be{redscalar}
\phi=\int^\sigma\frac{\sqrt{\frac{\M^2}{2}\paq{U+3\pa{\frac{\rd U}{\rd\sigma'}}^2}}}{U}d\sigma'
\ee
leading to the potential
\be{Wdef}
W(\phi)=\frac{\M^4}{4U^2}V(\sigma\pa{\phi}).
\ee
Let us note that these transformations can also be used when studying exact solutions for some cosmologies or static geometries which are
more complicated than FLRW flat universes (see e.g. \cite{we-non-Fried}).
 
In particular for the induced gravity case (\ref{induced}) and the potential (\ref{plpot}) one has
\be{EFinduced}
\tilde a=\frac{\sqrt{6\gamma}}{m_p}a\,\sigma,\;\phi=\frac{\Gamma}{6\gamma}m_p\ln\frac{\sigma}{\sigma_0},\; 
\ee
and
\be{indpot}
W(\phi)=\frac{m_p^4}{\pa{6\gamma}^2}\lambda\,M^{4-n}\sigma_0^{n-4}\exp\paq{\pa{n-4}\frac{6\gamma}{\Gamma}\frac{\phi}{m_p}},
\ee
where $m_p\equiv \sqrt{6}\M$. \\
While the equivalence of the two frames, the JF described by (\ref{lag0}) and the EF by (\ref{EFaction}), is expected at the classical level, at the quantum level it is not clear whether such an equivalence still holds. In a semiclassical context where the homogeneous background is treated classically while the inhomogeneous perturbations are quantised, at least in the linearized approximation, the frames are equivalent and, for example, the inflationary observables, such as the spectral indices, are the same.\\

On canonically quantising the classical system (\ref{EFaction}) in the minisuperspace approximation one finally obtains the following WdW equation
\begin{widetext}
\be{EFwdw}
\pag{\frac{1}{2m_p^2}\frac{\partial^2}{\partial\pa{\ln \tilde a}^2}-\frac{1}{2}\frac{\partial^2}{\partial \phi^2}+\frac{m_p^4\tilde a^6}{\pa{6\gamma}^2}\paq{\frac{\sigma_0}{M}\exp\pa{\frac{6\gamma}{\Gamma}\frac{\phi}{m_p}}}^{n-4}}\tilde\Psi\pa{\tilde a,\phi}=0,
\ee
\end{widetext}
where again a particular ordering has been chosen for the kinetic term associated with the scale factor $\tilde a$.\\
Let us now consider the EF transition, described by (\ref{EFinduced}) and (\ref{indpot}), after quantisation. The WdW equation for minisuperspace in the Jordan Frame is (\ref{WdW}) with (\ref{plpot}) and, on using the chain rule
\be{chain}
\frac{\partial}{\partial\pa{\ln \sigma}}=m_p{\frac{\Gamma}{6\gamma}}\frac{\partial}{\partial \phi}+\frac{\partial}{\partial\pa{\ln \tilde a}},\; \frac{\partial}{\partial\pa{\ln a}}=\frac{\partial}{\partial\pa{\ln \tilde a}}
\ee
one exactly finds equation (\ref{EFwdw}). We conclude that, at least in the minisuperspace approximation canonical quantisation and the transition from the JF to the EF (and vice versa) indeed commute. \\
As a consequence the exact solutions found in the JF can be mapped into exact solutions of (\ref{EFwdw}) in the EF.\\
Let us now consider, for simplicity, the solution obtained on starting from the ansatz (\ref{factor2}) with $n=4$. For such a case the scalar field potential is transformed into a cosmological constant $\rho_\Lambda=m_p^4/(6\gamma)^2$. In the Einstein frame the solution (\ref{factor2}) corresponds to
\be{efsol}
\tilde \Psi=\exp\paq{\mu\frac{6\gamma}{\Gamma}\frac{\phi}{m_p}}\bar \chi(\tilde a),
\ee
where $\bar \chi$ satisfies (\ref{wdwans2}). More general solutions can be obtained as a superposition of solutions of the form (\ref{efsol}). Let us note that, in the EF, the total homogeneous wave function can be factorised into the product of a wave function for the inflaton and that for the scale factor. It appears that for $\mu$ real the inflaton wave function is divergent at infinity in the EF while for $\mu$ imaginary it takes a plane wave form corresponding to an eigenstate of the (hermitean) field momentum operator with a real eigenvalue. Therefore the analysis of such a solution in the EF constrains $\mu$ to be imaginary.\\
The corresponding form for the gravitational wave function $\bar \chi$ is given by (\ref{eq-chi2b}) with $n=4$ and $\mu\equiv i\tilde \mu$ (with $\tilde \mu$ real). The solution found is a fully quantum solution with a well definited classical counterpart which can be obtained as follows. We first note that, at the classical level, the energy density of the inflaton fluid corresponding to a solution with a constant momentum $\pi_{\phi}$ is 
\be{rhocl1}
\rho_\phi=\frac{1}{2}\dot\phi^2=\frac{\pi_\phi^2}{2\tilde a^6}
\ee 
On considering a classical value for the momentum equal to its quantum eigenvalue $\pi_{\phi}=\frac{\tilde\mu}{m_p}\frac{6\gamma}{\Gamma}$ we finally obtain the corresponding classical Friedmann equation
\be{freq0}
H_{\rm cl}^2=\frac{2}{m_p^2}\pa{\rho_\phi+\rho_\Lambda}=\frac{m_p^2}{18\gamma^{2}}\pa{\frac{\tilde \mu^2\pa{6\gamma}^4}{2m_p^6\Gamma^2 \tilde a^6}+1}
\ee  
which has the attractor solution (\ref{clsol1}) in the $\tilde a\rightarrow \infty$ limit.\\
Let us now consider the modulus squared of the gravitational wave function $\bar \chi(\tilde a)$. If one, for simplicity, considers $c_1=1$ and $c_2=-i$) the wave function is
\be{Hwf}
\bar \chi\pa{\tilde a}= H_{2\mu\gamma/\Gamma}^{(2)}\pa{\frac{\sqrt{{2}}}{{18\gamma}}m_p^3\tilde a^3},
\ee
where $H_s^{(2)}(z)$ is a Hankel function of the second kind with an asymptotic behaviour ($z\rightarrow +\infty$) given by
\be{Hwfasy}
H_s^{(2)}(z)\simeq \sqrt{\frac{2}{\pi z}}\re^{-i\paq{z-\frac{\pi}{4}\pa{2s+1}}}\paq{1+i\frac{1-4s^2}{8 z}}.
\ee
Therefore, in the large $a$ limit and $\pi_\phi\gg m_p^{-1}$ the modulus squared of the gravitational wave function is
\be{modsq}
\bar \chi(\tilde a)^*\bar \chi(\tilde a)\simeq \frac{18\gamma\sqrt{2}}{\pi m_p^3\tilde a^3}\pa{1-\frac{\tilde\mu^2\pa{6\gamma}^4}{4\Gamma^2m_p^6\tilde a ^6}}\propto \frac{1}{\tilde a^3 H_{\rm cl}},
\ee
where the last relation holds for $a\, m_p\gg 1$. Thus the classical probability for a given classical solution is recovered and a well defined correspondence between the quantum and the classical solutions is established. Such a correspondence must hold both in the EF and in the JF and we can therefore conclude that the attractor given by (\ref{clsol1}) corresponds, at the quantum level, to a solution with $\mu=0$. The solutions with $\mu\neq 0$ describe the evolution during an approach towards the attractor.\\
Let us now consider the solution $\tilde \Psi$ given by (\ref{wavesol}) with $n=4$. In the EF 
\begin{eqnarray}
&&u_+=\frac{m_p^3\tilde a ^3}{\pa{6\gamma}^{3/2}}\pa{\re^\frac{3\phi}{m_p}+\re^{-\frac{3\phi}{m_p}}},\nonumber \\
&&v_+=\frac{m_p^3\tilde a ^3}{\pa{6\gamma}^{3/2}}\pa{\re^\frac{3\phi}{m_p}-\re^{-\frac{3\phi}{m_p}}}.
\label{uvEF}
\end{eqnarray}
Let us set, for example, $c_1=1$, $c_2=0$ and evaluate 
\begin{widetext}
\be{momsol2}
-i\frac{\partial}{\partial\phi}\tilde \Psi=3\frac{m_p^2\tilde a^3}{\pa{6\gamma}^{3/2}}\paq{\pa{q+\sqrt{q^2+B}}\re^{\frac{3\phi}{m_p}}+\pa{q-\sqrt{q^2+B}}\re^{-\frac{3\phi}{m_p}}}\tilde \Psi.
\ee
\end{widetext}
The expression on the r.h.s. of the above equality is the classical momentum as a function of $\phi$ multiplied by $\tilde \Psi$. Let us note that the general solution to the classical equations of motion for (\ref{EFaction}) with $W(\phi)=\Lambda$ leads to the following phase space trajectories:
\be{cltraj}
\dot \phi=\pm \sqrt{\frac{\Lambda}{2}}\paq{\tilde D\re^{\mp\frac{3\phi}{m_p}}-\tilde D^{-1}\re^{\pm\frac{3\phi}{m_p}}},
\ee
and correspondingly $\dot \phi\,\tilde a ^3=\pi_\phi={\rm const}$, which exactly reproduces the r.h.s. of (\ref{momsol2}) once the definitions of $B=\gamma/3$ and that of the cosmological constant in the EF ($\Lambda=m_p^4/\pa{6\gamma}^2$) are taken into account. Moreover one must identify 
\be{tD}
\tilde D=\frac{q}{\sqrt{B}}+\sqrt{\frac{q^2}{B}+1},
\ee
where the correspondence between the quantum eigenvalue $q$ and the classical integration constant has been shown. Let us note that the quantum solution (\ref{wavesol}) has a form similar to a ``generalised'' plane-wave and can be simultaneously associated with the general classical solution through the relation $\hat \pi_\phi \tilde \Psi=\pi_{\phi,\,\rm cl}\pa{\phi}\tilde\Psi$ (just as for the WKB case).\\

Therefore, in contrast with the case discussed for the solution (\ref{efsol}) which was connected to the classical solution on comparing the corresponding probability densities (in the large $\tilde a$ limit), in the EF, for the wave function (\ref{wavesol}), a correspondence between the quantum and the classical solutions is possible on examining the expression for the quantum momentum and its relation with its classical counterpart. Let us further note that, in this latter case, the quantum-classical correspondence based on the comparison of the probability density is not possible because of the plane-wave form of the solution (\ref{wavesol}) which trivially gives $\tilde\Psi^*\tilde \Psi=1$.\\

We further observe that the same classical solution can be obtained both starting from Eqs. (\ref{efsol}) or (\ref{wavesol}).

\section{Conclusions}
Inflation is currently believed to be the highest energy physics mechanism which can be tested by observations. Further, since it occurs for scales just a few orders of magnitudes below the Planck scale, it may be affected by the quantum gravitational effects. The correct description of gravity at such scales, when quantum effects become relevant, is not clear nonetheless it is reasonable to expect that the canonical procedure for the quantisation of gravity (and the resulting WdW equation) will lead to a sensible theory of quantum gravity, at least in the minisuperspace approximation, which can then be applied to the study of inflation. In this article our analysis focusses on the solution of the WdW equation for a set of induced gravity models, instead of just GR, and a minimally coupled inflaton. Induced gravity models are a natural generalisation of GR and, even if they were introduced many years ago, recently have become more and more attractive. When quantum effects become large enough a non minimal coupling to gravity naturally arises in the presence of a scalar field which then affects the observed Newton's constant. Higgs inflation belongs to this class of models since, in them, the scalar Higgs field is also responsible for inflation and generates the primordial inhomogeneities with spectra which are compatible with observations. Nowadays Higgs inflation (and the models related to it by a frame transformation) seems favoured by observations since it reconciles, within a common framework, Planck scale and Standard Model physics. Higgs inflation occurs during a phase which is dynamically indistinguishable from induced gravity with a nearly quartic potential and this motivates our quantum gravity approach to induced gravity.\\

The canonical quantisation of the inflaton-gravity system for the case of induced gravity leads to a WdW equation which can be solved exactly for power-law form potentials. The same class of models is exactly solvable classically as well and this fact, in principle, allows an exact comparison between the quantum solutions and their classical counterparts. Moreover it gives the opportunity of studying the equivalence between the JF and the EF description. Such an equivalence is defined classically but its extension to the (full) quantum level is not obvious. In this paper we found that, at least in the minisuperspace approximation, the EF and the JF are indeed equivalent in the sense that the frame transformation ``commutes'' with the canonical quantisation. We exploited such an equivalence in order to analyse the exact set of solutions found for the WdW equation. In particular we found two sets of one parameter independent solutions for it. These solutions have been transformed to the EF and their classical counterpart have been found. We have shown that the free parameter entering in the solutions has a classical counterpart and can be put in correspondence with the classical trajectories. Let us further observe that the EF version of the models studied corresponds to the power-law inflation case and is classically exactly solved, and we now also have its quantum counterpart.\\
  
The importance of the exact solution found is also reinforced by the possibility we now have of calculating the quantum gravitational effects on the primordial spectra generated during inflation following the approach already adopted for the minimally coupled case \cite{wdwpert}. The inclusion of inhomogeneities is certainly necessary in order to further clarify the much debated correspondence between the JF and the EF both theoretically and at the level of inflationary observables (primordial spectra).

\section{Acknowledgements}
Alexander Y. Kamenshchik is supported in part by the RFBR grant 17-02-01008.


\end{document}